\documentstyle[preprint,aps,pre]{revtex}
\begin{document}
\draft


\title{Enhancement of Stochastic Resonance in distributed systems due to a selective 
coupling}

\author{B. von Haeften$^1$, R. Deza$^1$\thanks{Electronic Address: deza@mdp.edu.ar} 
and H. S. Wio$^2$\thanks{Electronic Address: wio@cab.cnea.gov.ar,\newline 
http://www.cab.cnea.gov.ar/Cab/invbasica/FisEstad/estadis.htm}}

\address{(1) Departamento de F\'{\i}sica, Facultad de Ciencias Exactas y
Naturales, \\ Universidad Nacional de Mar del Plata, De\'an Funes 3350, 7600 
Mar del Plata, Argentina. \\
(2) Centro At\'omico Bariloche (CNEA) and Instituto Balseiro (CNEA and UNC)\\
8400 San Carlos de Bariloche, Argentina.}

\date{\today}

\maketitle

\begin{abstract}
Recent massive numerical simulations have shown that the response of a
``stochastic resonator'' is enhanced as a consequence of spatial
coupling.  Similar results have been analytically obtained in a 
reaction-diffusion model, using {\em nonequilibrium potential\/} 
techniques.  We now consider a field-dependent diffusivity and show 
that the {\em selectivity\/} of the coupling is more efficient for 
achieving stochastic-resonance enhancement than its overall value in 
the constant-diffusivity case.
\end{abstract}



The phenomenon of {\em stochastic resonance\/} (SR)---namely, the {\em
enhancement\/} of the output signal-to-noise ratio (SNR) caused by
injection of an optimal amount of noise into a periodically driven
nonlinear system---stands as one of the most puzzling and promising
cooperative effects arising from the interplay between {\em
deterministic\/} and {\em random\/} dynamics in a {\em nonlinear\/}
system.  The broad range of phenomena---indeed drawn from almost every
field in scientific endeavor---for which this mechanism can offer an
explanation has been put in evidence by many reviews and conference
proceedings, Ref.\cite{RMP} being the most recent and comprehensive
one, from which one can scan the state of the art.

Most phenomena that could possibly be explained by SR occur in {\em
extended\/} systems: for example, diverse experiments are being
carried out to explore the role of SR in sensory and
other biological functions \cite{biol} or in chemical systems
\cite{sch}.  Notwithstanding this fact, the overwhelming majority of
the studies made up to now are based on zero-dimensional systems,
while most of the features of this phenomenon that are peculiar to the
case of extended systems---or {\em stochastically resonating media\/}
(SRM)---still remain largely to be explored.  Particularly interesting
numerical simulations on arrays of coupled nonlinear oscillators have
been recently reported \cite{adi}, indicating that the {\it coupling}
between these stochastic resonators enhances the response of the
array, which exhibits moreover a higher degree of synchronization.
This effect has its counterpart in the continuum, as a study on the
overdamped continuous limit of a $\phi^4$ field theory shows
\cite{fabio}.  Recently---by exploiting the previous knowledge of the
{\em nonequilibrium potential\/} (NEP) \cite{GR} for a bistable
reaction-diffusion (RD) model \cite{I0}---one of us has shown 
analytically that the SNR increases with diffusivity in the range
explored \cite{wiocas}.

While considering a constant diffusion coefficient $D$ is a standard
approach, it is not the most general one: it is reasonable to expect
that the reported enhancement in the SNR by the effect of diffusion 
could depend in a more detailed way on $D$. In this regard, see for 
instance \cite{IncBul}. In this letter we consider the more realistic 
case of a field-dependent diffusion coefficient $D(\phi(x,t))$, and 
show that it causes an  {\em enhancement\/} of the SNR {\em still 
larger\/} than the one associated with a homogeneous increase of its 
amplitude.

The model under study---a one-dimensional, one-component RD model
describing a system that undergoes an electrothermal instability
\cite{I0}---can be regarded as the continuous limit of the coupled
system studied by Lindner {\em et al\/} \cite{adi}.  The field
$\phi(x,t)$ might describe the (time-dependent)
temperature profile in the ``hot-spot model'' of superconducting
microbridges \cite{I0}.  This model can be also regarded as a
piecewise-linear version of the space-dependent Schl\"ogl model for an
autocatalytic chemical reaction, and that for the ``ballast
resistor'', describing the so-called ``barretter effect'' \cite{RL}.
As a matter of fact, since in the ballast resistor the thermal
conductivity is a function of the energy density, the resulting
equation for the temperature field includes a temperature-dependent 
diffusion coefficient in a natural way  \cite{RL}.  Pointers to
other contexts in which a description containing a field-dependent
diffusivity becomes inescapable have been included in
Refs.\cite{strier,tsallis}.  

By adequate rescaling of the field, space-time variables and parameters, 
we get a dimensionless time-evolution equation for the field $\phi (x,t)$
\begin{equation}
\label{Ballast}
\partial_t\phi(x,t)=\partial_x\left(D(\phi)\partial_x\phi\right)+f(\phi)
\end{equation}
where $f(\phi)=-\phi+\theta(\phi-\phi_c)$, $\theta(x)$ is 
Heaviside's step function.  All the effects of the parameters that
keep the system away of equilibrium (such as the electric current
in the electrothermal devices or some external reactant concentration
in chemical models) are included in $\phi_c$. Moreover, since the
value of the field $\phi(x,t)$ corresponds in these models to
the {\em deviations\/} with respect to e.g.\ a reference temperature
$T_B>0$ (the temperature of the bath) in the ballast resistor or to a
reference concentration $\rho_0$ in the Schl\"ogl model, it is clear
that---up to a given {\em strict\/} limit (i.e.\ $\phi=-T_B$ for the 
ballast resistor)---some negative values of $\phi(x,t)$ are allowed.

As was done for the reaction term \cite{I0,wiocas}, a simple choice
that retains however the qualitative features of the system is to
consider the following dependence of the diffusion term on the field
variable
\begin{equation}
\label{diff}
D(\phi)~=~D_0(1+h\,\theta(\phi-\phi_c)),
\end{equation}
For simplicity, here we choose the same threshold $\phi _c$ for the
reaction term and the diffusion coefficient.  The more general
situation is left for a forthcoming work \cite{bernardo}.

We assume the system to be limited to a bounded domain $x \in [-L,L]$
with Dirichlet boundary conditions at both ends, i.e.\ $\phi(\pm
L,t)=0$.  The piecewise-linear approximation of the reaction term in
Eq.(\ref{Ballast})---which mimicks a cubic polynomial---was chosen
in order to find analytical expressions for its stationary
spatially-symmetric solutions.  In addition to the trivial solution
$\phi_0(x)=0$ (which is linearly stable and exists for the whole range
of parameters) we find another linearly stable nonhomogeneous
structure $\phi_s(x)$---presenting an excited central zone (where
$\phi_s(x)>\phi_c$) for $-x_c\le x\le x_c$---and a similar unstable
structure $\phi_u(x)$, which exhibits a smaller excited central zone.
The form of these patterns is analogous to what has been obtained in
previous related works \cite{I0}, as is shown in Fig. 1. The difference 
is that in the present case $d \phi/d x|_{x_c}$ is discontinuous and 
the area of the central zone depends on $h$. 

The indicated patterns are extrema of the NEP, which---among other
properties that we shall be using presently---is a Lyapunov
functional for the {\em deterministic\/} system introduced thus far.
In fact, the unstable pattern $\phi_u(x)$ is a {\em saddle-point\/} of
this functional, separating the {\em attractors\/} $\phi_0(x)$ and
$\phi_s(x)$ \cite{I0}.  The notion of a NEP has been thoroughly
studied, mainly by Graham and his collaborators \cite{GR}.  Loosely
speaking, it is an extension to non-equilibrium situations of the
familiar notion of (equilibrium) {\em thermodynamic potential\/}.  
For the case of a field-dependent diffusion coefficient $D(\phi(x,t))$ 
as described by Eq. (\ref{Ballast}), it reads \cite{I0,RL}
\begin{equation}
{\cal F}[\phi]=\int_{-L}^{+L}\left\{-\int_0^\phi D(\phi')f(\phi')\,d\phi'+
\frac{1}{2}\left(D(\phi)\frac{\partial\phi}{\partial x}\right)^2\,\right\}dx.
\end{equation}
Given that $\partial_t\phi~=~-(1/D(\phi)){\delta{\cal
F}}/{\delta\phi}$ one finds $\dot{{\cal F}}=-\int\left({\delta{\cal
F}}/{\delta\phi}\right)^2dx\leq 0$, thus warranting the
Lyapunov-functional property.  This NEP functional offers the
possibility of studying not just the linear but also the nonlinear---in 
the case at hand the {\em global\/}---stability of the patterns, following
its changes as the parameters of the model are varied \cite{I0}.

For a given threshold value $\phi _c^*$, both wells corresponding in
a representation of the NEP to the linearly stable states have the
same depth (i.e.\ both states are equally stable).  Figure 2 shows the
dependence of ${\cal F}[\phi]$ on the parameter $\phi _c$. As in previous 
cases, we analyze only the neighborhood of $\phi _c=\phi _c^*$ \cite{wiocas,wio2}.   
Here we also consider the neighborhood of $h =0$, where the main trends of the 
effect can be captured.

Now, with the aim of studying SR, we introduce a weak signal that
modulates the potential ${\cal F}$ around the situation in which the
two wells have the same depth.  This is accomplished by allowing the
parameter $\phi_c$ to oscillate around $\phi_c^{*}$:
$\phi_c(t)=\phi_c^{*}+\delta\phi_c\cos{(\Omega t+\varphi)}$, with
$\delta\phi_c \ll \phi_c(t)$.  We also introduce in Eq.(\ref{Ballast}) 
a fluctuating term $\xi(x,t)$, which we model (as is customary) as an 
additive Gaussian white-noise source with zero
mean value and a correlation function $\langle\xi(x,t)\,
\xi(x^{\prime},t^{\prime})\rangle=2\gamma\,\delta(t-t^{\prime})\,
\delta(x-x^{\prime}),$ thus yielding a stochastic partial differential
equation for the random field $\phi(x,t)$.  The parameter $\gamma$
denotes the noise strength \cite{com}.  

As in previous works \cite{I0}, we exploit a generalization to
extended systems of the Kramers-like result for the evaluation of the
decay time or ``mean-first-passage time" $\langle\tau\rangle$
\cite{fedo0,fedo}.  Here, those results are extended to the case of
field-dependent diffusivity, yielding \cite{bernardo}
\begin{equation}
\label{tau}
\langle\tau\rangle=\tau_0\,\exp\left\{\frac{{\cal W}[\phi,\phi_i]}
{2\gamma}\right\}.
\end{equation}
The functional ${\cal W}[\phi,\phi_i]$ ($\phi_i$ indicates the initial
metastable state, which at each instant may be either $\phi_0$ or
$\phi_s$), that is the solution of a Hamilton-Jacobi-like equation, has 
the following expression
\begin{eqnarray}
\label{WW}
{\cal W}[\phi,\phi_i]=\int_{-L}^{+L}\,dx\,\left\{\left(\frac{D(\phi)}{2}
\left(\frac{\partial\phi}{\partial x}\right)^2- U(\phi)\right)
\,- \left(\frac{D(\phi_i)}{2}\left(\frac{\partial\phi_i}
{\partial x}\right)^2-U(\phi_i)\right)\right\},
\end{eqnarray}
with $U(\phi)=\int_0^{\phi}d\phi' f(\phi')$.  The prefactor $\tau_0$
in Eq.(\ref{tau}) is essentially determined by the curvature of the
NEP ${\cal F}[\phi,\phi_c]$ at its extrema.

The calculation of the SNR proceeds, for the spatially extended
problem, through the evaluation of the space-time correlation function
$\langle\phi(y,t)\phi(y^{\prime},t^{\prime})\rangle$.  To do that we
use a simplified point of view, based on the two-state approach \cite{mnw}, 
which allows us to apply some known results almost directly.  To proceed 
with the calculation of the correlation function, we need to evaluate the transition
probabilities $W_{\pm}\propto \langle\tau\rangle^{-1}$, which appear
in the associated master equation.  For small $\delta\phi_c,$
\[{\cal W}[\phi,\phi_i]\approx{\cal W}[\phi,\phi_i]_{\phi^*_c}+
\delta\phi_c\left(\frac{\partial{\cal W}[\phi,\phi_i]}{\partial\phi_c}
\right)_{\phi_c^{*}}\cos{(\Omega t+\varphi)}.\]
Solving such a master equation up to first order in $\delta\phi_c$ it
is possible to evaluate the correlation function.  Its double Fourier
transform, the generalized susceptibility $S(\kappa,\omega)$,
factorizes in this approach, and the relevant term becomes a function
of $\omega$ only (the corresponding expressions are omitted, see 
\cite{wio2} for details).  

It is worth noting that many of the results exposed here (e.g.\ the 
profiles of the stationary patterns and the corresponding values of the 
NEP) are exact.  The only approximations involved in the
calculation of the SNR are the standard ones, namely the Kramers-like
expression in Eq.(\ref{tau}) and the two-level approximation used for
the evaluation of the correlation function \cite{mnw}.

Using the definition from Ref.\cite{mnw} for the SNR at the excitation
frequency (here indicated by $R$), the final result is
\begin{equation}
\label{snr}
R\sim(\frac{\Lambda}{\tau_0\gamma})^2
\exp(-{\cal W}[\phi,\phi_i]_{\phi^*_c}/\gamma),
\end{equation}
where $\Lambda=(d{\cal W}[\phi,\phi_i]/d\phi_c)_{\phi_c^{*}}\delta\phi_c$,
and $\tau_0$ is given by the asymptotically dominant linear stability
eigenvalues:
$\tau_0=2\pi(\left|\lambda^{un}\right|\overline{\lambda^{st}})^{-1/2}$
($\lambda^{un}$ is the unstable eigenvalue around $\phi_u$ and
$\overline{\lambda^{st}}$ is the average of the smallest eigenvalues
around $\phi_0$ and $\phi_s$).  Equation (\ref{snr}) is analogous to
the results in zero-dimensional systems, but here $\Lambda$,
$\tau_0$ and ${\cal W}[\phi,\phi_i]_{\phi^*_c}$ contain all the
information regarding the spatially extended character of the system.

In Fig.3 we depict the dependence of $R$ on the noise intensity
$\gamma $, for several (positive) values of $h$.  These curves show
the typical maximum that has become the fingerprint of the stochastic
resonance phenomenon.  Figure 4 is a plot of the value $R_{max}$ of
these maxima as a function of $h$.  The dramatic increase of
$R_{max}$, of several $dB$ for a {\em small} positive variation of $h$,
is apparent and shows the strong effect that the selective coupling 
(or field-dependent diffusivity) has on the response of the system.

The present prediction prompts to devise experiments (for instance,
through electronic setups) as well as numerical simulations taking
into account the indicated selective coupling.  
This result could be of relevance for technological applications such as 
signal detection and image recognition, as well as for the solution of 
some puzzles in biology (mammalian sensory systems, ionic channels in 
cells).  

The present form of analysis is being extended to (the bistable regime 
of) multicomponent models of the activator-inhibitor type since---in
addition to their applications to systems of chemical (e.g.\
Bonhoffer-Van der Pol model) and biological (e.g.\ FitzHugh-Nagumo
model) origins---these models are related to spatio-temporal
synchronization problems\cite{RMP,biol,adi}.  An effective treatment
of models of this type gives rise to a non-local coupling, which would
compete with the nearest neighbour coupling $D(\phi)$ presented here
\cite{wio2}.\\ \\

{\bf Acknowledgments:} The authors thank J.I.Deza for his help with
the numerical calculations and preparation of the figures and 
V. Grunfeld for a revision of the manuscript. HSW and
RD thank the ICTP for the kind hospitality extended to them during
their stay.  Partial support from the Argentine agencies CONICET
(grant PIP 4953/97) and ANPCyT (grant 03-00000-00988), is also
acknowledged.

\newpage 

\begin{figure}[tbp]
\caption{$\phi_s(y)$ is the {\em stable\/} pattern, and $\phi_u(y)$ the 
{\em unstable\/} one (similar in form but exhibiting a smaller excited
central zone); they are extrema of the NEP: $\phi_u(y)$ is a {\em
saddle-point\/} of this functional, separating the {\em attractors\/}
$\phi_0(y)$ and $\phi_s(y)$ [7]. Here we indicate $\phi_0(y)$ and $\phi_s(y)$
with a full line, $\phi_u(y)$ with a dotted line, while the value of $\phi_c$
is shown with a dashed line.}
\end{figure}

\begin{figure}[tbp]
\caption{Dependence on $\phi_c$ of the value of $F$ calculated for the
patterns in Fig.1.  The stable branch $F_s$ (indicated with a full line) 
and the unstable one $F_u$ (indicated with a dotted line) collapse at a 
critical value of $\phi_c$. At a certain value $\phi_c=\phi_c^*$ (indicated 
by an arrow), the value of $F_s$ becomes positive and $\phi_s(x)$
becomes metastable.}
\end{figure}

\begin{figure}[tbp]
\caption{SNR $R$ as a function of the noise intensity $\gamma$ (Eq.(6)), for 
three values of $h$: $h=0.0$ (full line), $h=-0.25$ (dashed line) and $h=0.25$ 
(dotted line).  We have fixed $L=1$, $D_0=1$, $\delta\phi_c=0.01$ and $\Omega=0.01$.}
\end{figure}

\begin{figure}[tbp]
\caption{Maximum $R_{max}$ of the SNR curve (Fig.3) as a function of $h$, 
for three values of $D_0$: $D_0=0.9$ (dashed line), $D_0=1.$ (full line) and 
$D_0=1.1$ (dotted line). The arrows {\bf a} and {\bf b} indicate 
the response gain due to an homogeneous increase of the coupling and to 
a selective one respectively. The larger gain in the second case is apparent.
The inset shows the dependence of
$R_{max}$ on $D_0$ for $h=-0.25$ (lower line), $h=0$ and $h=0.25$
(upper line).}
\end{figure}


\begin{references}
\bibitem{com} Strictly speaking---since there must be
an absorbing boundary at the lower limit (i.e.\ $\phi=-T_B$ for the case
of the ballast resistor) where the noise has to vanish---the
fluctuations cannot be Gaussian.  However, our experience (supported
by the qualitative agreement found in \cite{wiocas} with the
simulations in \cite{adi}) indicates that the differences introduced
by the Gaussian negative tail do not affect the results.
\end{references}
\end{document}